\documentclass[11pt,twoside]{article}

\usepackage{asp2004}
\usepackage{epsf}
\usepackage{psfig}
\usepackage{lscape}

\begin{document}
\title{Radio-mm-FIR Photometric Redshifts for (sub-)mm Galaxies}   %%% Fill in title
\author{Itziar Aretxaga$^1$, David H. Hughes$^1$, James S. Dunlop$^2$}   %%% Fill in author names
\affil{$^1$Instituto Nacional de Astrof\'{\i}sica, \'Optica y Electr\'onica\\ 
Aptdo. Postal 51 y 216, 72000 Puebla, Mexico.\\    
$^2$Institute for Astronomy, University of Edinburgh, Royal Observatory,
Edinburgh, EH9 3HJ, UK.
}    
%%% Fill in author affiliations

\begin{abstract} %%% Abstract to run on from here.
We present a comparison between the published optical, IR and CO
spectroscopic redshifts of 86 (sub-)mm galaxies and their photometric
redshifts as derived from long-wavelength (radio--mm--FIR) photometric
data. The redshift accuracy measured for 13 sub-mm galaxies with at
least one robustly-determined colour in the radio--mm--FIR regime and
additional constraining upper limits is $\Delta z \approx 0.30$.  
This accuracy degrades to $\Delta z
\approx 0.65$ when only the 1.4GHz/850$\mu$m spectral index is used, 
as derived from the analysis of a sub-sample of 58 galaxies with
robustly determined redshifts.  Despite the wide range of spectral
energy distributions in the local galaxies that are used in an
un-biased manner as templates, this analysis demonstrates that
photometric redshifts can be efficiently derived for sub-mm galaxies
with a precision of $\Delta z < 0.5$ using only the rest-frame FIR to
radio wavelength data, sufficient to guide the tuning of broad-band
heterodyne observations (e.g. 100m GBT, 50m LMT, ALMA) or aid their
determination in the case of a single line detection by these
experiments.
\end{abstract}

%%% MAIN BODY OF TEXT GOES HERE. CONSULT "INSTRUCTIONS FOR AUTHORS USING
%%% LATEX2E MARKUP", SECTIONS 2.3-2.6 FOR HELP WITH EQUATIONS, FIGURES,
%%% AND TABLES.

\section{Introduction}   %%% Top level section head (remove "%" symbol)

The next generation of wide-area extragalactic submillimetre and
millimetre surveys, will produce large samples ($\sim 10^{3} -
10^{5}$) of distant, luminous starburst galaxies.  The dramatic
increase in the number of detected galaxies makes it necessary to
identify a selection technique that can efficiently generate redshift
selected subsamples that are then explored in greater detail with
optical/IR and mm follow-up facilities, such as ALMA.

Given the underlying assumption that we are witnessing high rates of
star formation in these (sub-)millimetre galaxies (see Blain, in this
volume), then we expect them to have the characteristic FIR peak and
steep submillimetre (Rayleigh-Jeans) spectrum which is dominated by
thermal emission from dust grains heated to temperatures in the range
$\sim 20-70$\,K by heavily obscured young, massive stars.  The
observed radio--FIR luminosity correlation in local starburst galaxies
(e.g. Helou, Soifer
\& Rowan-Robinson  1985), that links the radio synchrotron emission
from supernova remnants with the later stages of massive star
formation, is also expected to apply to these galaxies.

Thus, in recent years, a considerable amount of effort has been
invested in assessing the accuracy with which these broad continuum
features in the spectral energy distributions (SEDs) of submillimetre
galaxies, at rest-frame mid-IR to radio wavelengths can be used to
provide photometric-redshifts (Hughes et al. 1998; Carilli \& Yun
1999, 2000; Dunne, Clements \& Eales 2000; Rengarajan \& Takeuchi
2001; Yun \& Carilli 2002; Hughes et al. 2002, Aretxaga et al. 2003,
2005; Wiklind 2003; Blain, Barnard \& Chapman 2003; Hunt \& Maiolino
2005; Laurent et al. 2006).

This paper is an update on previous work (Aretxaga et al. 2005) that
assessed the accuracy of the FIR-mm-radio photometric redshift
techniques given the larger samples of spectroscopic redshifts
recently published in the literature (see Chapman and Tacconi, in this
volume, and references therein, for the primary sources of comparison
data).

The cosmological parameters adopted throughout this paper are
$H_0=67$~km\,s$^{-1}$\,Mpc$^{-1}$, $\Omega_{\rm M}=0.3$,
$\Omega_{\Lambda}=0.7$.

\section{Accuracy of the 1.4GHz/850$\mu$m spectral index as a redshift 
diagnostic}   %%% Top level section head (remove "%" symbol)

%\subsection{}   %%% Second level section head (remove "%" symbol)
%\subsubsection{}   %%% Lowest level section head (remove "%" symbol)
%\section*{}    %%% Unnumbered top level section head (remove "%" symbol)
%\subsection*{}   %%% Unnumbered second level section head (remove "%" symbol)

The 1.4GHz/850$\mu$m spectral index is discussed following 3 different
prescriptions: the one-template maximum-likelihood technique
originally designed by Carilli \& Yun (1999, 2000), denoted as $z_{\rm
phot}^{\rm CY}$; a maximum likelihood technique which simultaneously
fits the 20 local templates of starbursts, ULIRGs and AGN used by
Aretxaga et al. (2003, 2005), denoted as $z_{\rm phot}^{\rm A}$; and a
multi-template maximum-likelihood technique that uses the 3 blue
compact dwarf SEDs of Hunt \& Maiolino (2005) that reproduce the
radio--mm--FIR data of sub-mm galaxies (the templates for NGC\,5253,
and two flavours of II\,Zw\,40 SEDs with radio slopes $\alpha=-0.5$
and $-0.1$), which we denote as $z_{\rm phot}^{\rm HM}$. For all three
techniques the error bars of the photometric redshifts were derived by
bootstrapping on the reported photometric and calibration errors and,
in the case of $z_{\rm phot}^{\rm CY}$, the error estimated by Carilli
\& Yun (2000), to allow for a difference in templates, is added in
quadrature to the errors derived by bootstrapping the photometry.

\begin{figure}
\plotfiddle{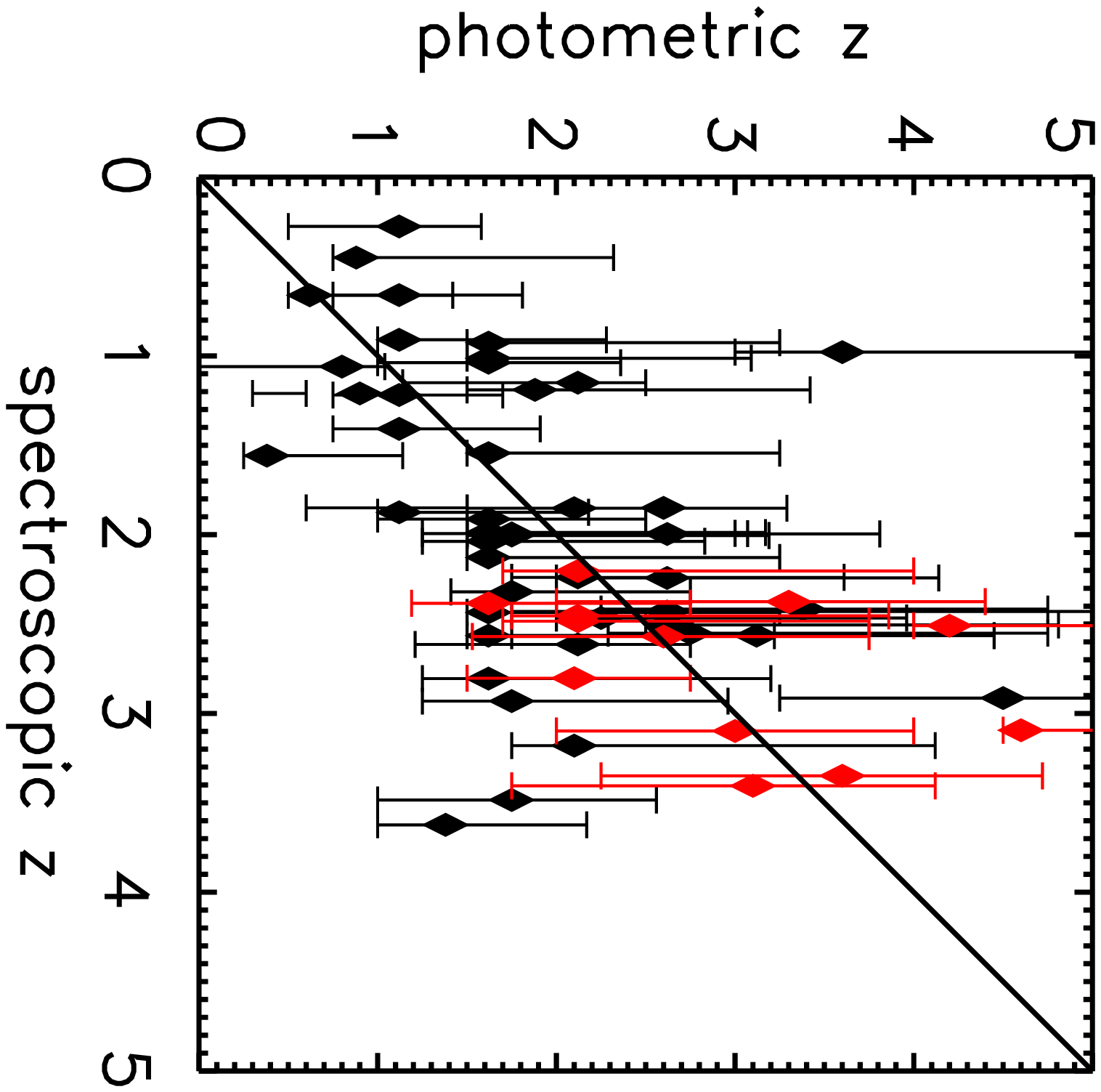}{7cm}{90}{50}{50}{200}{-25}
\plotfiddle{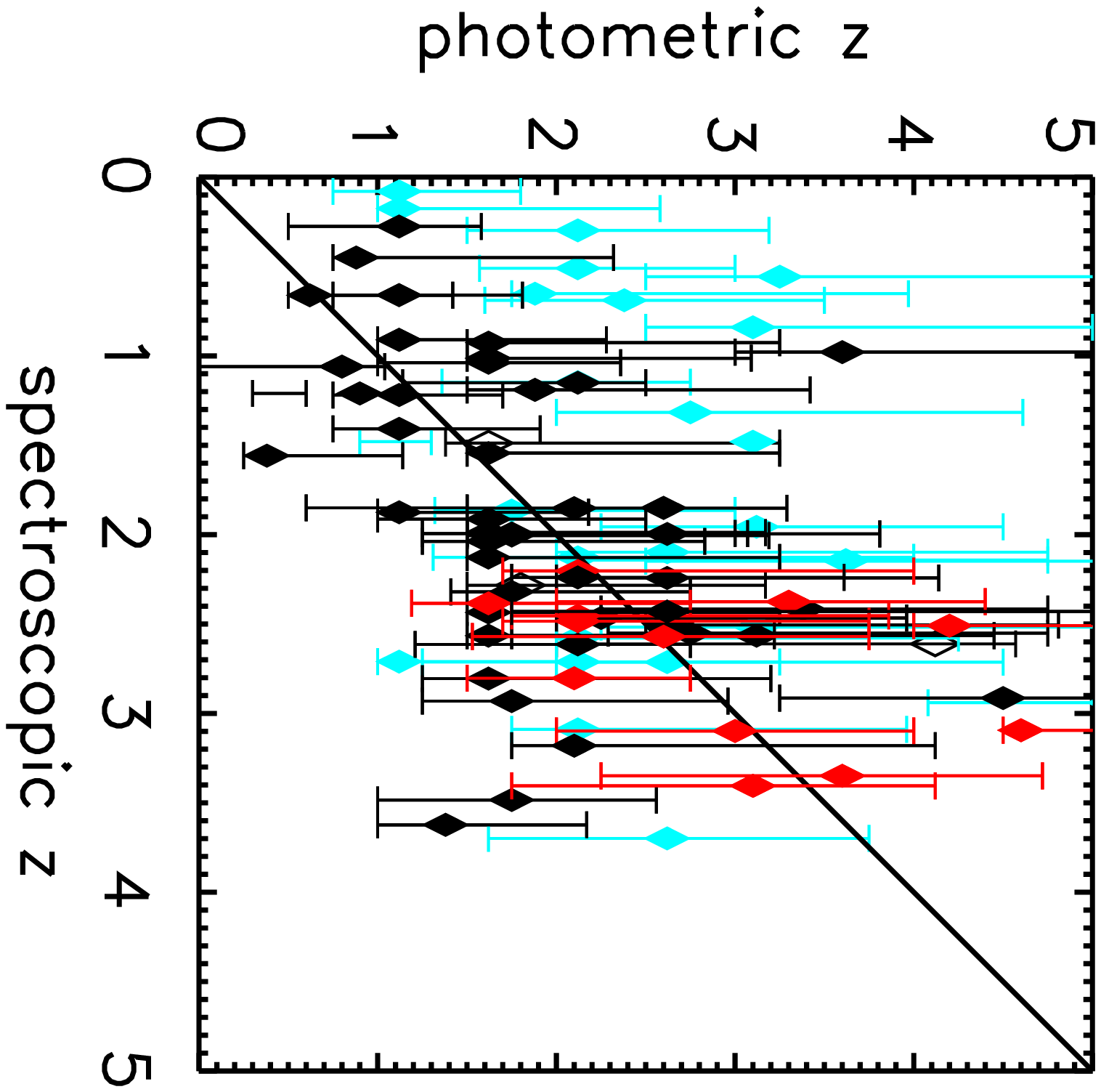}{7cm}{90}{50}{50}{200}{-40}
\caption{Comparison of spectroscopic and photometric redshifts
derived from the 1.4GHz/850$\mu$m spectral index for a sample of 86
galaxies. {\bf (Top)} Robust sample of undisputed radio/optical/IR
counterpart associations of the sub-mm galaxies, and spectroscopic
redshifts derived from 2 or more lines: in dark grey, CO confirmed
optical/IR spectroscopic redshifts; and in black, redshifts based on
optical/IR spectra with two or more characteristic emission or
atmospheric/winds absorption lines.  The error bars represent 68\%
confidence intervals in the determination of the redshift, and the
diamonds the mode of the distributions. {\bf (Bottom)} The same for
all 86 galaxies, where empty symbols represent optical/IR redshifts
that have failed to obtain confirmation with CO observations, and
light grey symbols are redshifts that have been disputed in the
literature or that are based on only one characteristic line, and thus
are not considered robust enough to be included in the sample above.
}
\label{fig:CY_A}
\end{figure}

We have assembled a sample of 86 objects which have
published optical/IR or CO spectroscopic redshifts ($z_{\rm spec}$)
and accompanying
photometry to assess the accuracy of the techniques described in this
paper.

Using the 1.4GHz to 850$\mu$m color ratio and the $z^{\rm A}_{\rm
phot}$ prescription, we find a mean accuracy of $z^{\rm A}_{\rm
phot}-z_{\rm spec}\sim 0.65$ (Fig.~\ref{fig:CY_A}) for a robust
sub-sample of objects with undisputed interpretation about their
optical/IR/radio counterparts and spectroscopic redshifts derived from
the identification of more than two spectral features. This subsample
does not include powerful radio-loud AGN, for which the templates used
in the photometric redshift analysis are not appropriate.  Restricting
the analysis only to those galaxies with CO spectroscopic redshifts,
the measured accuracy is $z^{\rm A}_{\rm phot}-z_{\rm spec}\sim 0.6$.
The error distribution for the sample of galaxies is centered at
$-0.15$, which we do not regard as significant as it is well below the
resolution used in the Monte Carlo simulations that aid the
calculation of the photometric redshifts. The precision degrades as
the redshift increases, as expected from the 1.4GHz/850$\mu$m 
spectral index, which flattens beyond $z=3$ (Carilli \& Yun 2000), 
leading to a measured $z^{\rm
A}_{\rm phot}-z_{\rm spec}\sim 1.0$ at $3\le z
\le 4$.
Using all objects with published photometry, regardless of whether the
spectroscopic redshift is suspected to be incorrect or not, or is not regarded as robustly determined, the
overall accuracy over the $0\le z \le 4$ regime degrades to $\Delta z
= 0.8$.

We find that for the same robust sample of objects, $z^{\rm CY}_{\rm
phot}$ has systematically larger errors, $\Delta z \sim 0.9$, while
the $z^{\rm HM}_{\rm phot}$ provides a similar $\Delta z
\sim 0.7$ dispersion compared to the results of $z^{\rm A}_{\rm phot}$.  The
$z^{\rm HM}_{\rm phot}$ prescription gives, however, poorer
estimations of the redshift for the CO confirmed sources ($\Delta
z\sim 0.85$) than the $z^{\rm A}_{\rm phot}$ technique ($\Delta z \sim
0.6$).

\section{Fitting the full rest-frame radio-FIR SED}   %%% Top level section head (remove "%" symbol)

For a few tens of sources, multiwavelength data in the radio to FIR
regime is available, beyond the 1.4GHz and 850$\mu$m photometry. These
additional data can also be used to place extra constraints on the
redshift of the sources by sampling the FIR bump. This information has
been exploited by several groups, including our own, using a wide
array of fitting techniques and SEDs (e.g. Yun \& Carilli 2002,
Wiklind 2003, Laurent et al. 2006). Our own particular technique is
based on Monte-Carlo simulations to also take into account
constraining prior information such as the number counts of submm
galaxies, the favoured luminosity/density evolution up to $z\approx
2$, and the amplification or clustering of a certain field (Hughes et
al. 2002, Aretxaga et al. 2003, 2005).  We only offer a brief summary
of the technique here.  We generate a catalogue of 60$\mu$m
luminosities and redshifts for mock galaxies from an evolutionary
model for the $60\mu$m luminosity function that fits the observed
850$\mu$m number-counts.  Randomly-selected template SEDs are drawn
from a library of local starbursts, ULIRGs and AGN, to provide
FIR--radio fluxes.  The SEDs cover a wide-range of FIR luminosities
($9.0 < {\rm log} L_{\rm FIR}/L_{\odot} < 12.3$) and temperatures
($25<T/K<65$). The fluxes of the mock galaxies include both
photometric and calibration errors, consistent with the quality of the
observational data for the sub-mm galaxy detected in a particular
survey. We reject from the catalogue those mock galaxies that do not
respect the detection thresholds and upper-limits of the particular
sub-mm galaxy under analysis.  The redshift probability distribution
of a sub-mm galaxy is then calculated as the normalized distribution
of the redshifts of the mock galaxies in the reduced catalogue,
weighted by the likelihood of identifying the colours and fluxes of
each mock galaxy with those of the sub-mm galaxy in question.

\begin{figure}[t]
%\plottwo{LH850.1_prob.ps}{LH850.1_SED_z2.4.ps}
\plotfiddle{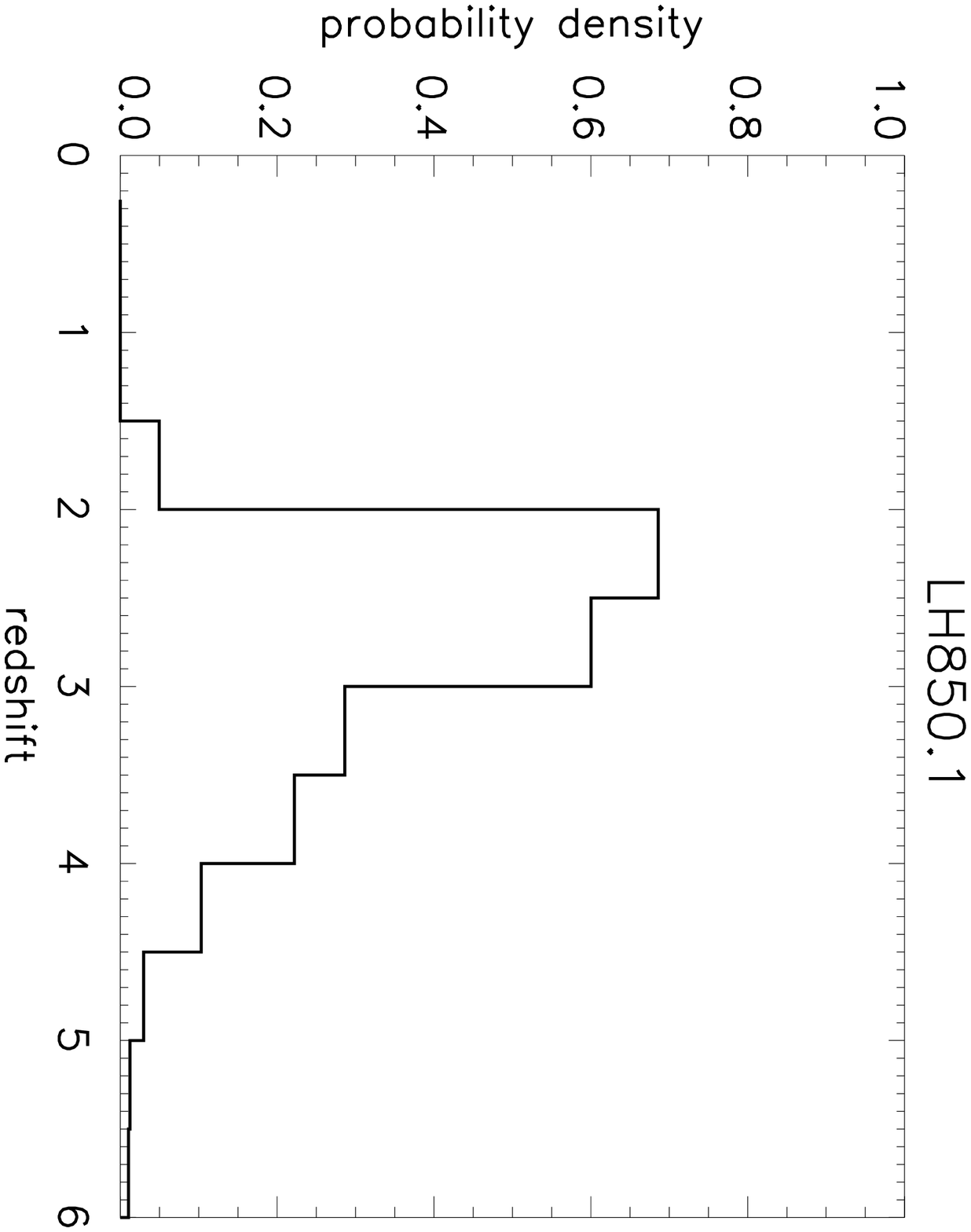}{5cm}{90}{30}{30}{15}{-20}
\plotfiddle{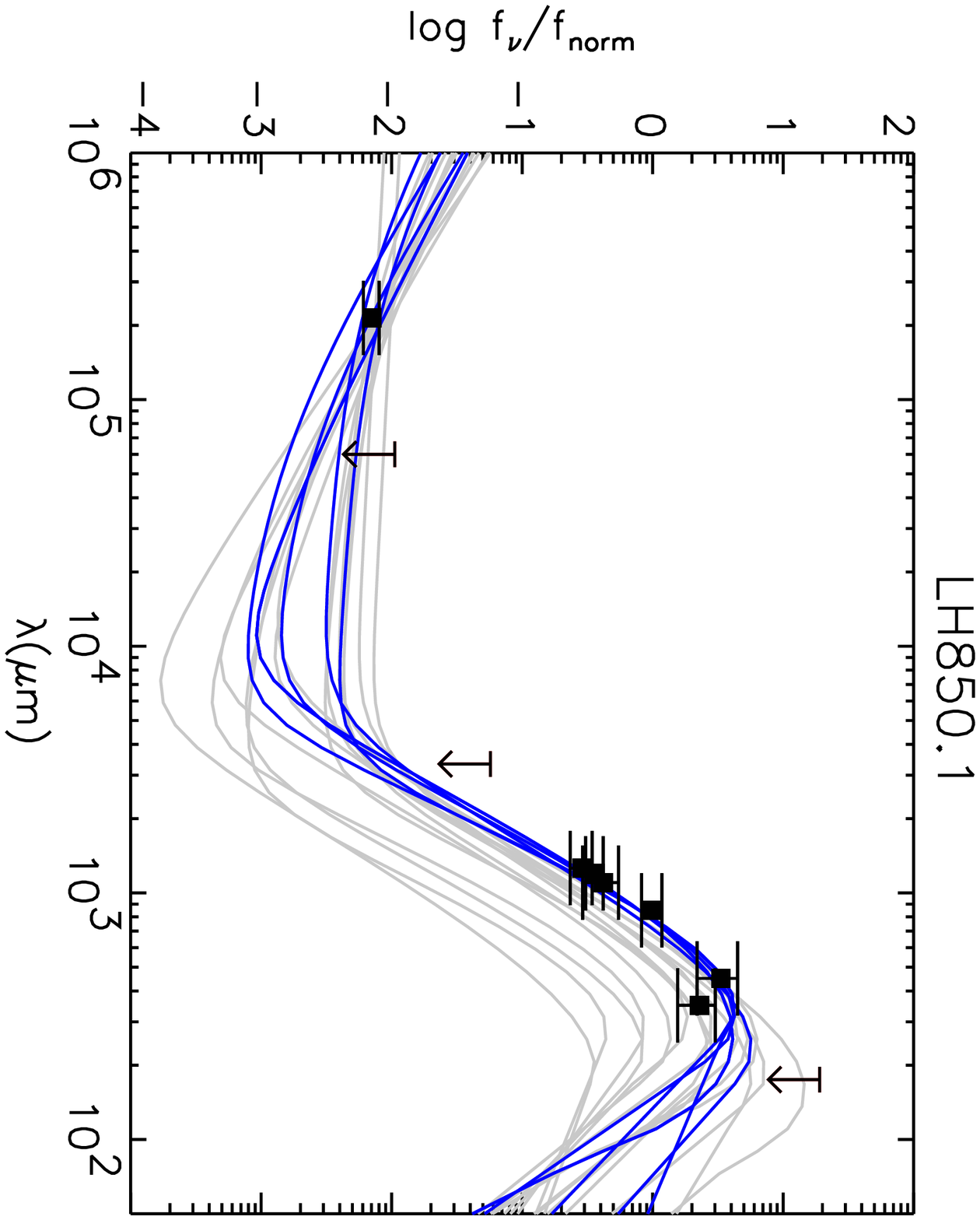}{5cm}{90}{30}{30}{200}{135}
\vspace*{-5.5cm}
\caption{
{\bf Left} Redshift probability density distribution derived for
LH850.1. The probability density peaks at $z\approx 2.4$ (as seen in
higher resolution simulations). {\bf Right} The observed SED of
LH850.1 normalised to the flux density at 850$\mu$m is shown as
squares (detections) and arrows (3$\sigma$ upper limits).  The
template SEDs (lines) are redshifted to the spectroscopic redshift
$z=2.15$ and scaled in flux to fit the points and arrows through
survival analysis. The template SEDs {\em at this redshift},
compatible within the $3\sigma$ error bars with the SED of LH850.1,
are displayed as darker lines.  A fit at $z = 2.4$ also selects 5
compatible SEDs with the photometry of the object, yielding a figure
that is almost indistinguishable from this one by eye.  }
\label{fig:LH850.1}
\end{figure}

Fig.~2 shows an example of our method for LH850.1, the brightest
source of the SCUBA 8mJy survey in Lockman Hole field (Scott et
al. 2002, Ivison et al. 2002, Greve et al. 2004, Laurent et al. 2005,
2006), that tentatively has been assigned the optical/IR spectroscopic
redshift of a plausible counterpart at $z_{\rm spec}=2.15$ (Chapman et
al. 2005, Ivison et al. 2005).

\begin{figure}
\plotfiddle{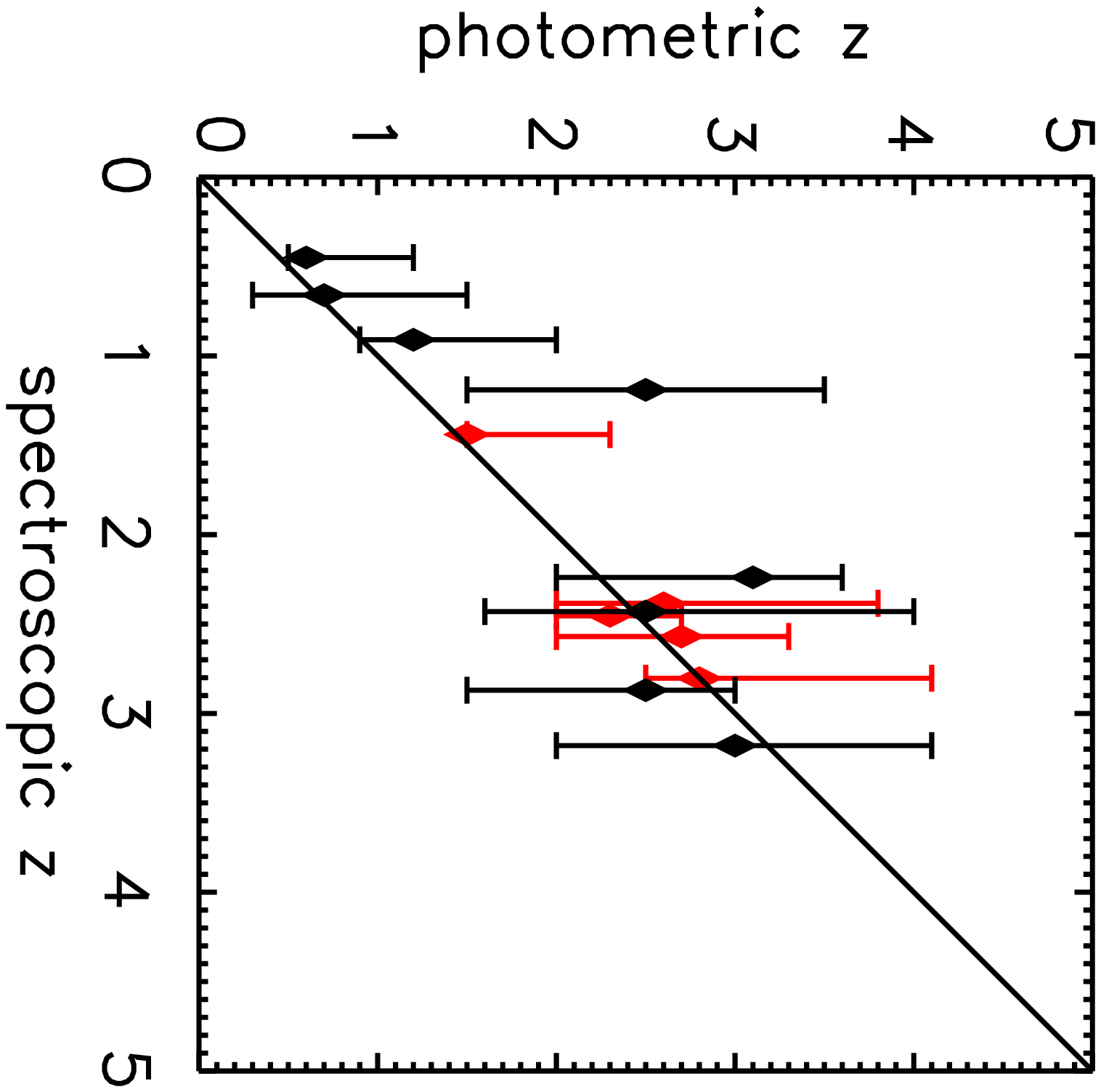}{7cm}{90}{50}{50}{200}{-20}
\plotfiddle{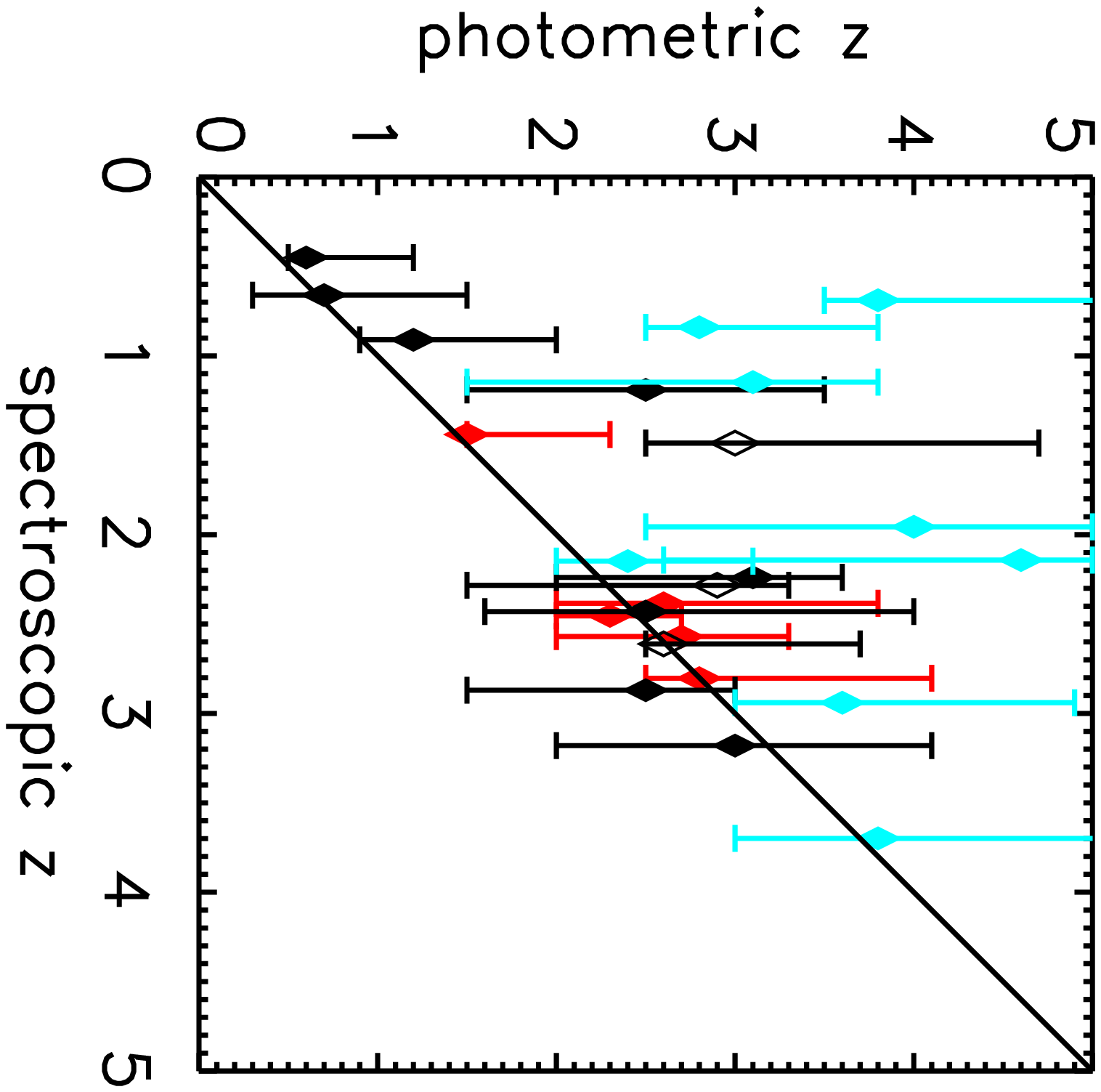}{7cm}{90}{50}{50}{200}{-40}
\caption{Comparison of spectroscopic and photometric redshifts
derived from the full radio-FIR SED for a sample of 24 galaxies with
accompanying multiwavelength photometry. {\bf (Top)} Robust sample of
undisputed radio/optical/IR counterpart association of the sub-mm
galaxies, and spectroscopic redshifts derived from 2 or more lines: in
dark grey, CO confirmed optical/IR spectroscopic redshifts; and in
black, redshifts based on optical/IR spectra with two or more
characteristic absorption or emission lines. {\bf (Bottom)} The same
for all 24 galaxies, where empty symbols represent optical/IR
redshifts that have failed confirmation in CO, and light grey symbols
correspond to galaxies with redshifts that have been disputed in the
literature or that are based on one characteristic line only, and thus
are not considered robust enough to be included in the sample above.
}
\label{fig:SED_A}
\end{figure}

Using the full SED information of a robust sub-sample of 13 objects
(Fig.~\ref{fig:SED_A}) with undisputed identification of their
optical/IR/radio counterparts and spectroscopic redshifts derived from
the measurement of two spectral features or more, we find a mean
accuracy of $z^{\rm A}_{\rm phot}-z_{\rm spec}\sim 0.3$.  Restricting
the analysis only to those 5 galaxies with CO spectroscopic redshifts,
the measured accuracy is $z^{\rm A}_{\rm phot}-z_{\rm spec}\sim 0.1$.
The error distribution for the robust sample of galaxies is centered
in $\Delta z=+0.17$, which is within the resolution afforded by the
Monte Carlo simulations that aid the photometric redshifts. In
contrast with the simple 1.4GHz/850$\mu$m spectral index, the
precision does not degrade as the redshift increases. Three extra
galaxies can be included in the same analysis, that have not been
confirmed by a CO detection at the published optical/IR spectroscopic
redshift. This could be due to an erroneous redshift or insufficient
sensitivity in the search for a CO line. Including these 3 objects in
the analysis, we find an overall precision of $\Delta z=0.3$.  Using
all objects with published photometry, regardless of whether the
spectroscopic redshift is questioned or not, the overall accuracy over
the $0\le z \le 4$ regime degrades to $\Delta z = 0.65$, with some
significant outliers (lower panel of Fig. 3).

The two outliers at the lowest redshift correspond to LH850.8 (at
$z_{\rm spec}=0.689$) and N2850.1 (at $z_{\rm spec}=0.840$), and their
observed SEDs cannot be reproduced by any of the SED templates in our
library at the published spectroscopic redshift (Fig.~4, see also
Aretxaga et al. 2005 and Laurent et al. 2006), or by the blue compact
dwarf SEDs of Hunt \& Maiolino, which have been claimed to provide
better fits to the overall sub-mm galaxy photometry than those of
giant star-forming galaxies --- although the inspection of Fig.4,
shows that these three templates lie within the range of the templates
of ULIRGs, AGN and local star-forming galaxies used in Aretxaga et al
(2003).  The robustness of the optical associations for N2850.1 and
LH850.8 has been questioned in the literature (Chapman et al. 2002,
Ivison et al. 2005). Apart from a true association, it also has been
suggested that a plausible low-$z$ lens magnifies the true higher-$z$
sub-mm galaxy, or that the true counterpart is identified by an
alternative radio source.  The CO-confirmation of the low-redshift
nature of these galaxies would be an important evidence for the
existence of a different starburst SED within the sub-mm galaxy
population that is not seen in the local analogs, and which should,
therefore, be incorporated into the photometric redshift
estimations. For the time being we consider them in the tentative
category of spectroscopic redshifts.

\begin{figure}[t]
%\plottwo{LH850.1_prob.ps}{LH850.1_SED_z2.4.ps}
\plotfiddle{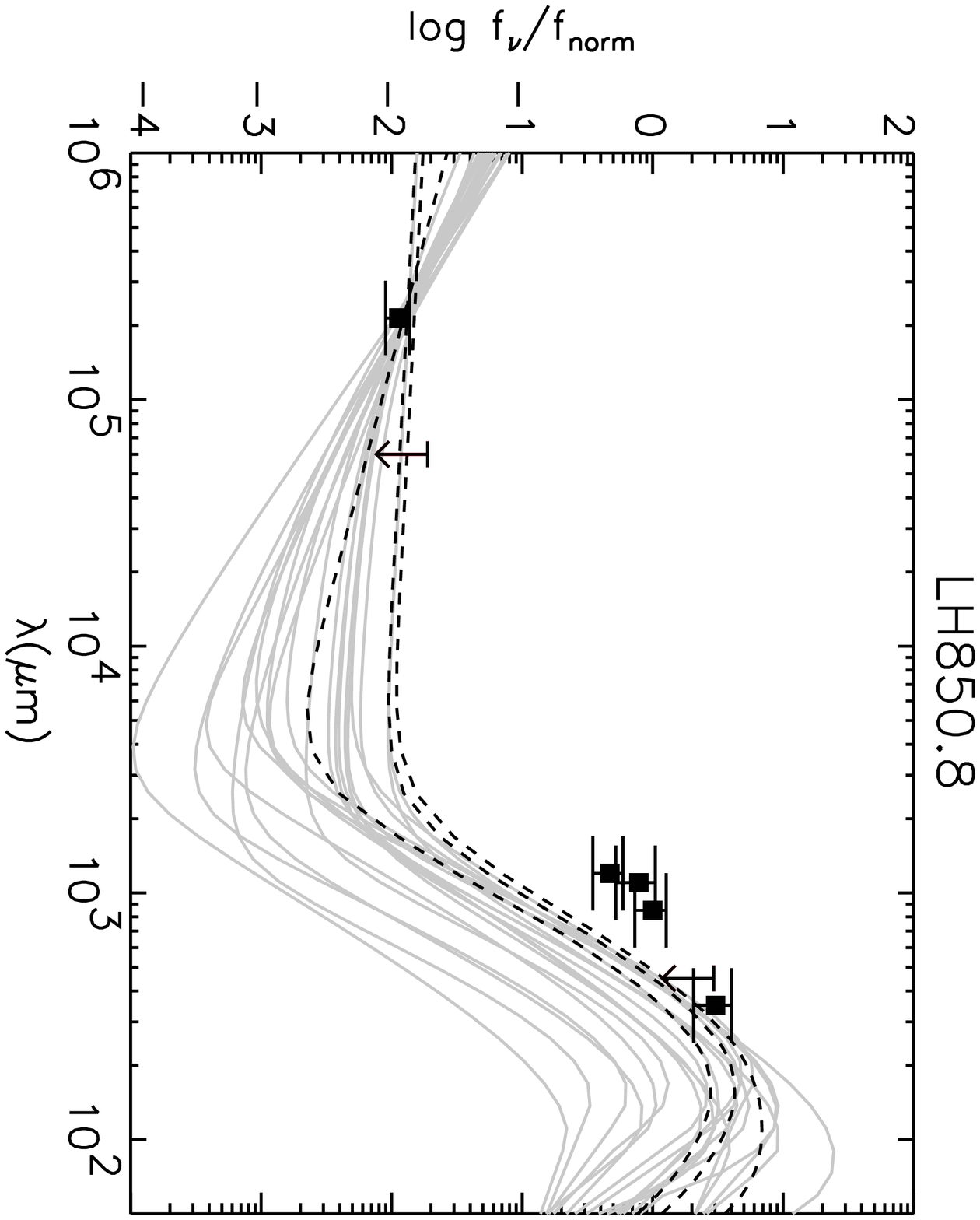}{5cm}{90}{30}{30}{15}{-20}
\plotfiddle{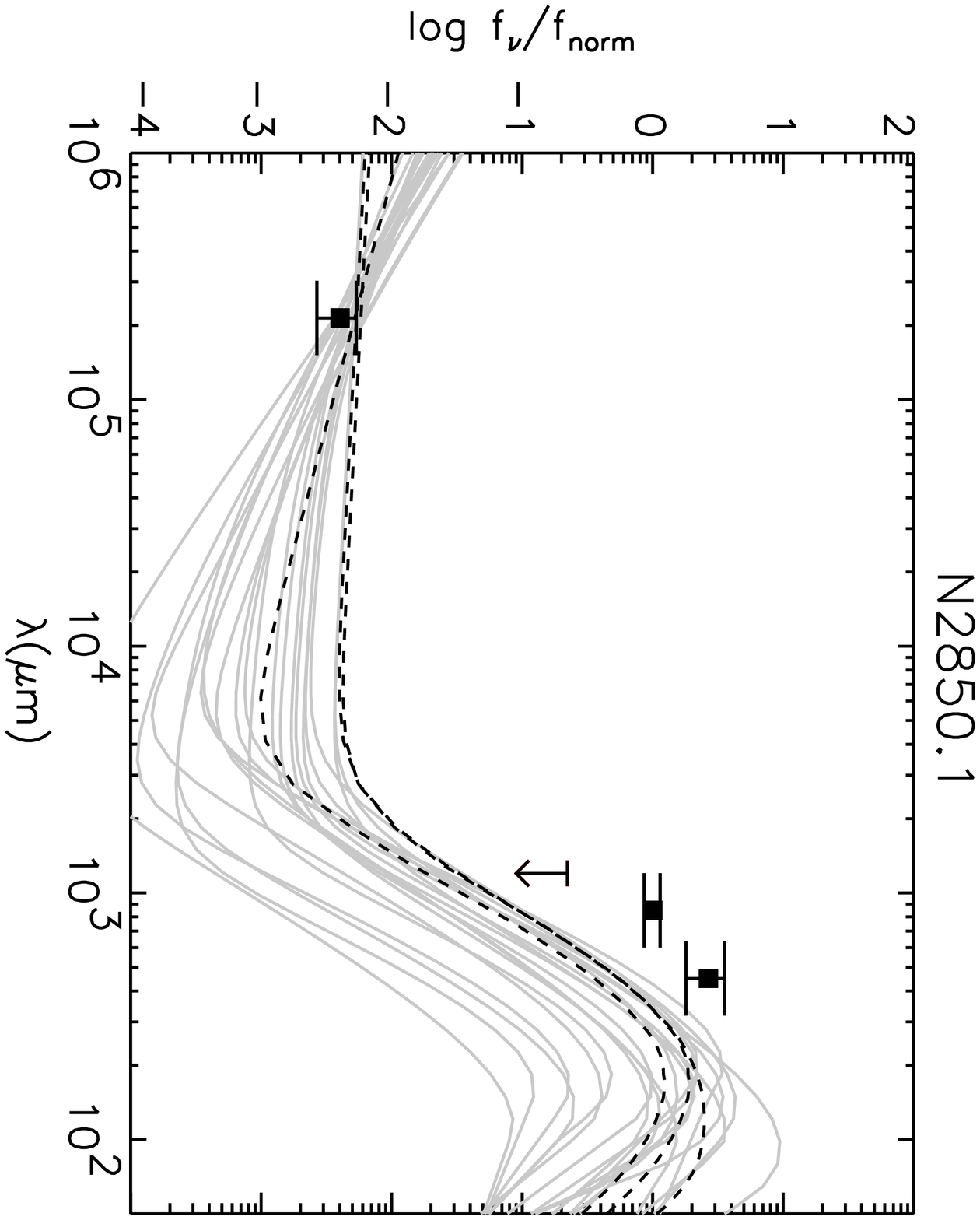}{5cm}{90}{30}{30}{200}{135}
\vspace*{-5.5cm}
\caption{
 The observed SEDs of LH850.8 and N2850.1, normalised to the flux
 density at 850$\mu$m is shown as squares (detections) and arrows
 (3$\sigma$ upper limits).  The template SEDs (lines) are redshifted
 to their respective spectroscopic redshifts $z_{\rm spec}=0.689$ and
 $z_{\rm spec}=0.840$ (Chapman et al. 2005) and scaled in flux to fit
 the points and arrows through survival analysis. There are no
 template SEDs {\em at these redshifts}, compatible within the
 $3\sigma$ error bars of these objects. In light grey we represent the
 20 templates of Aretxaga et al. (2003) and in dashed lines the 3
 templates of Hunt \& Maiolino (2005).}
\label{fig:catastrophic}
\end{figure}

\section{Prospects}   %%% Top level section head (remove "%" symbol)

%In the next decade, following the commissioning and subsequent
%operation .... of a wide-variety of new facilities (both large-format
%continuum cameras and telescopes) operating at mid-IR to millimetre
%and radio wavelengths (e.g. SCUBA-2, LaBOCA, LMT, ASTRO-F, Herschel,
%Planck, ACT, SPT, e-VLA, ... others ....), it will be possible to
%coordinate the new generation of sensitive cosmological surveys over
%a significant volume of the Universe.
%
%We can therefore anticipate extremely large flux-limited catalogues
%($\sim 10^3 - 10^6$ sources) of various populations of dust-enshrouded
%extragalactic sources, distributed over all cosmic epochs, that collectively 
%generate the entire FIR--mm extragalactic background. These new FIR--radio
%wavelength catalogues will then be used to determine the history of
%obscured starformation in the Universe and complement those studies of
%galaxy evolution at optical and IR wavelengths.

%As discussed in this paper, 
The combination of more sensitive
rest-frame radio to FIR data with more complete wavelength coverage,
that define the SEDs of individual galaxies identified in the next
generation of FIR--radio surveys, will derive photometric-redshifts
with even greater accuracy than those presented here. Furthermore,
given the depth and area of these future surveys it will
be extremely difficult, and often impossible, to provide useful
follow-up data at optical and IR wavelengths. Thus we expect the use
of photometric-redshifts to
quickly increase in importance as they influence the interpretation of
these new cosmological surveys.

The additional power of photometric redshifts however is their ability
to efficiently select from these new FIR-radio catalogues, sub-samples
of sources in well-defined redshift intervals that can be targeted by
the next generation of broad-band (sub)millimetre spectroscopic
receivers presented at this meeting. 
The future photometric redshifts from radio--FIR data
will be essential in maximising the use of telescope time on
facilities such as ALMA by limiting the selection of receiver tunings
required to successfully search for and detect redshifted CO-lines.
Since two CO-lines, identified with specific transitions, are required
to unambiguously determine a spectroscopic redshift, then the combined
use of a single CO-line and the photometric-redshift distribution can
place important constraints on the possible frequencies that must
search for a confirming second-line (see Yun in these proceedings).
The ideal solution is to combine a telescope with sufficient
collecting area, to provide the sensitivity, and a millimetre
spectrometer with sufficient bandwidth to always detect two CO-lines,
or alternatively a CO-line and CI, regardless of the redshift of the
source. The Large Millimetre Telescope (LMT) and its Redshift Receiver
(Erickson -- these proceedings) almost satisfy these
requirements with the exception of the redshift interval ($ 0.5< z <
1$ and $ z \approx 2.1 $ (Hughes et al. - these proceedings).

To conclude, the different technological approaches described in this
meeting to provide sensitive broad-band receivers at
millimetre-wavelengths, with stable and flat spectral-baselines, will
have an enormous impact on the follow-up of the most heavily-obscured
continuum sources for which optical and IR techniques fail. The secure
individual spectroscopic redshifts, and redshift distribution for the
populations of heavily-obscured starburst galaxies, that will be
derived from molecular-line emission at (sub)millimetre wavelengths,
will provide critical data to understand the biases inherent in the
identification of the counterparts that ultimately lead to the
optical--IR spectroscopic redshifts. 
%A more exciting prospect,
%however, is that 
These CO-line data will determine
the general physical properties of the gas in the molecular ISM of galaxies in
the high-redshift Universe and provide estimates of their dynamical
masses. 
%Furthermore the same data can provide dynamical evidence for
%hierarchical galaxy formation at the earliest cosmological epochs via
%galaxy-interactions and merging which stimulates these luminous
%episodes of massive star-formation.

\acknowledgements %%% Text of acknowledgements runs on after this command.
IA and DHH gratefully acknowledge support from CONACYT grants 39548-F
and 39953-F to conduct their research, and from NRAO and the
conference organizers to participate in this vibrant workshop.

%%% THE BIBLIOGRAPHY
%%%
%%% CONSULT SECTION 3 OF "INSTRUCTIONS FOR AUTHORS" FOR HOW TO USE NATBIB.
%%% AUTHORS ARE ENCOURAGED TO USE EITHER THE "THEBIBLIOGRAPY" ENVIRONMENT
%%% BY UNCOMMENTING (DELETING THE "%" SYMBOL) THE COMMANDS BELOW, OR BY
%%% USING THE BIBTEX ENVIRONMENT. TO FIND OUT WHICH IS APPLICABLE TO YOUR
%%% CONTRIBUTION, CONSULT THE VOLUME EDITORS FOR YOUR PROCEEDINGS.
%%%

%\newpage

\end{document}